# The component groups structure of DPPC bilayers obtained by specular neutron reflectometry

Michal Belička[a,b*], Yuri Gerelli[c], Norbert Kučerka[b,d], Giovanna Fragneto[c]

[a] Faculty of Mathematics, Physics and Informatics, Comenius University, SK-842 48 Bratislava, Slovakia

[b] Faculty of Pharmacy, Comenius University, SK-832 32 Bratislava, Slovakia

[c] Institut Laue-Langevin, FR-38042 Grenoble Cedex 9, France

[d] Frank Laboratory of Neutron Physics, Joint Institute of Nuclear Research, Dubna, Moscow Region, 141700, Russia

---

[*] Corresponding author at: Department of Physical Chemistry of Drugs, Faculty of Pharmacy, Comenius University, Odbojárov 10, SK-832 32 Bratislava, Slovakia
Tel.: +421 2 50117 281; fax: +421 2 50117 100.

Email address: belicka@fpharm.uniba.sk (M. Belička)

## Keywords:

Specular neutron reflectometry
Lipid bilayer
Scattering density profile model
Supported bilayer
Floating bilayer
Dibehenoylphosphocholine
Dipalmitoylphosphocholine

# Abstract

Specular neutron reflectometry (SNR) was measured on a system of a floating bilayer consisting of 1,2-dipalmitoyl-d62-*sn*-glycero-3-phosphocholine (d62-diC16:0PC) deposited over a 1,2-dibehenoyl-*sn*-glycero-3-phosphocholine (diC22:0PC) bilayer at 25 and 55 °C. The internal structure of lipid bilayers was described by a one-dimensional scattering length density profile (SLDP) model, originally developed for the evaluation of small-angle scattering data. The corresponding model reflectivity curves successfully describe the experimental reflectivity curves of a supported bilayer in the gel phase and a system of a floating bilayer in the liquid-crystalline phase. The reflectivity data from the supported bilayer were evaluated individually and served further as an input by the data treatment of floating bilayer reflectivity curves. The results yield internal structure of a deposited and floating bilayer on the level of component groups of lipid molecules. The obtained structure of the floating d62-diC16:0PC bilayer displays high resemblance to the bilayer structure in the form of unilamellar vesicles, however, simultaneously it shows rate of fluctuations in comparison to unilamellar vesicle bilayers.

# 1. Introduction

Biological lipid membranes represent natural boundaries in living cells, where they define intra- and intercellular compartments containing specific biochemical environments. Hydrocarbon chains located in the centre of biomembranes form a hydrophobic barrier with different rates of penetration for water, ions and small molecules. Simultaneously, they serve as docking platforms for various biological macromolecules and influence their functions. All physico-chemical properties of biomembranes are altered by the bilayer composition and surrounding medium properties (temperature, pressure, composition, etc.). In principle, they are the result of complicated lipid-lipid, lipid-environment and lipid-macromolecule interactions (Heimburg, 2007). Molecular arrangement inside lipid biomembranes at an equilibrated state reflects these interactions. Therefore, each experiment technique capable of probing the biomembrane internal structure provides a very valuable tool for the study of such complex dynamical systems like biomembranes. This is the case of specular neutron reflectometry (SNR) (Daillant et al., 2005).

A wide variety of lipids and other molecules present in natural biological membranes makes structural studies difficult due to their complicated composition, therefore artificial lipid bilayers with a controlled composition are preferred as simpler biomembrane models. The form of lipid bilayers used also depends also on requirements of a given experimental technique. Uni- and multilamellar vesicles are primarily used by neutron or X-ray diffraction and small-angle scattering experiments and can be easily prepared in several ways (Hope et al., 1986; MacDonald et al., 1991; Nieh et al., 2009). Main drawbacks of lipid vesicles systems are their tendency to aggregate into structures with higher complexity and polydispersity of radii, which complicates data treatment. Moreover, the obtained bilayer profile information is extracted from the averaged scattering over all space configurations of the bilayer. The stacks of aligned fully hydrated bilayers are used as well. They are usually prepared from dried deposited organic solvent-lipid solutions brought into contact with water after solvent removal (Katsaras et al., 1992; Tristram-Nagle et al., 2002). The stacks prepared in this way consist of hundreds of bilayers, hence, when considered as one-dimensional crystals, they are excellent structures primarily for diffraction techniques. Their disadvantages are that only averaged bilayer structure can be obtained and that it is not possible to investigate the interactions of larger macromolecules (e.g. polypeptides, nucleic acids, proteins) with lipid bilayers due to the need to study them in hydrated environments as they are not stable in bulk water. These complications can be solved by application of SNR on a single supported bilayer (Sackmann, 1996; Gutberlet et al., 2004; Vacklin et al., 2005) or a floating bilayer over a supported one (Fragneto et al., 2001; Daillant et al., 2005; Fragneto et al., 2012) or a chemically grafted bilayer (Hughes et al., 2008) at the solid/water interface. The surface of the solid substrate (usually silicon) can be hydrophilic (e.g. with a $SiO_2$ layer) or hydrophobic (e.g. with an Au layer). Whereas a membrane interacts with the solid surface, hence its fluidity and fluctuations may not be in a biologically relevant state, the floating bilayer interacts with the supported one in the same way as bilayers in multilamellar vesicles or in stacks, and thus it can be considered as a more suitable biomembrane model.

SNR has proved to be a powerful and reliable method for the structure investigation of nanometer-scale films located at media interfaces (e.g. silicon/water, water/oil, water/air). In principle, it utilizes the wave properties of neutrons and the fact they are reflected and refracted on the common boundary of two media with different scattering length densities (SLD). Cold neutrons, produced by nuclear reactors or spallation sources, are exceptionally well suited for these studies as they deeply penetrate samples without their destruction, in contrast to X-rays. In combination with the well-known contrast variation technique (Heller, 2010) they allow to detect structures at the level of the Ångstrom. In a SNR experiment the measured reflectivity, $R$, is given by the ratio of the reflected beam to the incident beam intensities, reflectivity $R$ and it captures the (averaged) internal profile of an irradiated planar structure. By SNR only the structure in the direction of transferred momentum is considered, i.e. in the direction of bilayer normal. The inhomogeneities in the structure plane are averaged.

As for many other scattering techniques the lack of phase information in the reflected beam involves the application of a model to obtain the SLD profile. Models used for the neutron reflectometry data evaluation are usually directly based on the so-called box model (Daillant et al., 2005), in which the profile of a deposited structure is represented by strips (boxes) with constant SLDs, parameterized by their thicknesses, SLDs and the roughnesses of their borders. Particularly, in the case of lipid bilayers whole regions of the same nature like hydrocarbon chains, polar heads with intercalated water molecules or water molecules between a supported and a floating bilayer are represented by distinct homogeneous layers. The construction of a corresponding model reflectivity curve follows from basic optical principles (Born and Wolf, 1999).

The main aim of our work is to investigate the possibilities of a more detailed bilayer model for application in SNR studies focused on lipid bilayers. Both, a supported and a floating bilayer are represented by the scattering density profile (SDP) model, based on a model proposed by Wiener and White (1992) and recently successfully applied for small-angle X-ray and neutron scattering experiments (Kučerka et al., 2008, 2009, 2011a). Similar model has already been applied by Shekhar et al. (2011) for the evaluation of SNR on supported lipid membranes. As a system of a floating bilayer consists of two, in general different, bilayers, we divided the whole process of data treatment into two main parts. In the first step we studied only a supported bilayer. In the next step we applied the obtained structure of the supported bilayer as an input for a more complicated floating bilayer system. The floating bilayer was measured at two different temperatures to compare its structure in a gel and a fluid phase.

## 2. Materials and methods

*2.1 Samples preparation*

1,2-dibehenoyl-*sn*-glycero-3-phosphocholine (diC22:0PC) and 1,2-dipalmitoyl-d62-*sn*-glycero-3-phosphocholine (d62-diC16:0PC) were purchased from Avanti Polar Lipids (Alabaster, USA). All organic solvents were used as received from Sigma Aldrich (St Louis, USA). Fresh Milli-Q water (18 MΩ cm, named $H_2O$ in the following) and $D_2O$ of isotopic 99 % purity were supplied by the Institut Laue-Langevin (ILL), Grenoble. All water was degassed prior to use to avoid formation of air bubbles in the solid-liquid cells during high temperature measurements. Silicon substrates in the form of $8 \times 5 \times 2$ cm$^3$ blocks with a single polished side were used as solid support for depositions. Shortly before sample preparation, the block was cleaned in chloroform, acetone and ethanol subsequently in an ultrasonic bath in each solvent for 15 minutes. Afterwards the block was exposed to ozone for 30 min and rinsed with $H_2O$.

The process of sample preparation was carried out in a Nima 1212D Langmuir trough (Nima Technology, Coventry, UK) filled with $H_2O$ and cooled down to 13 °C. Monolayers at the water/air interface were prepared from chloroform lipid solutions at 1 mg/ml concentration, which were spread in small droplets on a water surface. After evaporation of the organic solvent (15 min), lipid monolayers were slowly compressed up to a lateral pressure of 40 mN m$^{-1}$. Langmuir-Blodgett monolayer transferts were then performed at a constant pressure. In order to prepare stable floating bilayer systems diC22:0PC molecules were used to prepare the bilayer facing the solid support. On its top a floating bilayer composed by diC16:0PC molecules was deposited. The whole process of the floating bilayer system preparation comprised of a combination of the Langmuir-Blodgett (vertical) and the Langmuir-Schaefer (horizontal) deposition techniques, as described in detail elsewhere (Fragneto et al., 2001). After the last deposition the silicon block was sealed in standard solid-liquid cell (provided by ILL) without removing it from the water in order to avoid the contact of the deposited sample with air. The holder was equipped with two valves for solvent exchange and cooling system for keeping the sample at desired temperature. The sample was prepared 24 hours before measurements and was stored at 8 °C in the cold room. A simpler adhered bilayer composed of diC22:0PC molecules was prepared and characterized as a reference for data analysis.

*2.2 Measurements*

SNR measurements were performed at the Institut Laue-Langevin (Grenoble, France) at the high flux D17 reflectometer (Cubitt and Fragneto, 2002). The instrument was operated in time-of-flight (ToF) mode using an interval of neutron wavelengths between 2 and 18 Å and two incident angles 0.8° and 3.2°. The covered $q$ range (where $q = \sqrt{\frac{4\pi}{\lambda}\sin(\theta)}$ is the momentum transfer and $\theta$ is scattering angle) ranged from 0.005 to 0.2 Å$^{-1}$. All the samples were measured in three different H$_2$O/D$_2$O mixtures (having different SLD) and at 25 °C and 55 °C (below the main phase transition of diC22:0PC and respectively below and above this transition in diC16:0PC (Marsh, 2013)). The direct experimental data obtained from the reflectometer were treated using the COSMOS routine of LAMP software package (Richard et al., 1996), through which they were converted into $R(q)$ curves. Data files generated by COSMOS contain information about the experimental $q$-resolution and this was used during data analysis.

*2.3 Data analysis*

*Reflectometry principles*

The specular neutron reflectivity $R(\vec{q})$, where $\vec{q}$ is transferred momentum, is defined as the ratio between the intensities of a reflected and an incident beams, under conditions that a reflected beam lies in the incident plane and the angles of reflectance and incidence are the same. It follows from this specification that 1) the direction of transferred momentum is orthogonal to the interface, and 2) structures lying in the plane perpendicular to $\vec{q}$ are averaged and consequently only this averaged structure along the bilayer normal has an impact on $R(q)$.

Neutrons enter a silicon block through a side and reach the Si/SiO$_2$ surface with deposited material. The SiO$_2$ layer with adhered lipid bilayers and water environment can be approximated as a system of layers (boxes) parallel with the Si/SiO$_2$ surface with constant SLDs. As it follows from the Schrödinger equation for the given geometry, each layer can be characterized by a refractive index

$$n = \sqrt{1 - \frac{\lambda^2}{\pi}\rho}, \quad (1)$$

where $\lambda$ is the wavelength of incident neutrons and $\rho$ is the SLD of layer. For systems composed by weakly absorbing nuclei, as it is for the samples investigated in the present paper, absorption can be neglected. At each interface a neutron wave is partly reflected back (into the same layer) and partly refracted or transmitted into a neighbour layer depending on their refractive indexes. Hence, including multiple reflections of neutron waves, models based on layered structures become fully dynamical. The final reflected beam from the whole system of layers can be calculated using basic optical principles (Born and Wolf, 1999). In the current work we utilize the Parratt recursion relation (Parratt, 1954), in which the final reflectivity $R(q)$ of a stratified medium is calculated by a series of relations

$$\begin{aligned} R(q) &= \|X_N\|^2, \\ X_j &= e^{-i2k_j z_j} \frac{r_{j,j-1} + X_{j-1} e^{i2k_{j-1} z_j}}{1 + r_{j,j-1} \cdot X_{j-1} e^{i2k_{j-1} z_j}}, X_0 = 0, \\ r_{j,j-1} &= \frac{k_j - k_{j-1}}{k_j + k_{j-1}}, \\ k_j &= \sqrt{q^2/4 - 4\pi(\rho_j - \rho_{Si})}, \end{aligned} \quad (2)$$

where $j$ indexes layer interfaces from the last one marked with $j = 0$, below which is only water solvent, to $j = N$ which labels the sample's entering interface at the silicon block. $z_j$ is the position of $j^{th}$ interface, $k_j$ is the normal component of a wave vector in $j^{th}$ layer with SLD $\rho_j$ and $\rho_{Si}$ is the SLD of a silicon block. In principle, the partitioning of the SLD profile along the bilayer normal into sections representing layers and the application of the Parratt relation can be done in two ways. The first way is to associate layers with significant structures, like the whole component groups, on the normal of investigated bilayers, to calculate their SLD and consequently to apply Parratt's procedure to compute the corresponding theoretical $R(q)$. The second way is the choice of any preferred SLD model profile, like the SDP model already mentioned above, its division into appropriately small parts, inside of which SLD can be considered as constant, and the application of Parratt's procedure. We chose the second way to apply a more detailed model and to directly compare the results to those already published in literature on similar lipid systems (Kučerka et al., 2008, 2009, 2011b).

*Scattering density profile*

In fact as a model for the SLD profile of bilayer we used the component density profile the SDP model, which is based on the model originally proposed by Wiener and White (1992) and later elaborated by Kučerka et al. (2008, 2009, 2011) for application in scattering experiments as the SDP model.

In this model a lipid molecule is divided into several component groups represented by their probability distributions along the bilayer normal. The components are: choline methyl groups, phosphate+$CH_2CH_2N$, carbonyl-glycerol (GC), hydrocarbon methylene groups ($CH_2$) and methyl groups of hydrocarbon chains ($CH_3$). In the current work we slightly modified the SDP model and described the whole choline group (choline methyl groups+$CH_2CH_2N$, CHO) and phosphate (PO) as a standalone component groups. The SLD profile of bilayer is constructed through the so-called *primitive cell* of a bilayer, i.e. in analogy with crystallography, the simplest repetitive region of a space forming the whole pattern. In our case the primitive cell is a volume containing a single lipid molecule in the form of a cylinder with bases in the centre and on the outer side of a bilayer. The area of its base corresponds to an important structure parameter - *interface area per lipid molecule A*. Each individual component inside the primitive cell is represented by its volumetric distribution $v_i(z)$, which describes the distribution of its volume along the bilayer normal. One can easily construct the SLD profile of a bilayer from the component volume distributions, while considering water molecules as an ideally filling background as in (Kučerka et al., 2008). The SLD profile $\rho_{norm}(z)$ can be written as

$$\rho_{norm}(z) = \left(1 - \sum_{j \neq CH_3} v_j(z)\right)\rho_W + \sum_{j \neq CH_3} v_j(z)\rho_j + v_{CH_3}(\rho_{CH_3} - \rho_{CH_2}). \tag{3}$$

A special attention is paid to the methyl groups. They are located inside one layer together with methylene groups and therefore, following the same complementary principle as applied to other components and water background, they replace "methylene background".

Volumetric distributions of components $v_i(z)$ are defined through the probability distributions of components $p_i(z)$, component volumes $V_i(z)$ and $A$. Probability distributions of methyl, carbonyl-glycerol, phosphate and choline groups are represented by Gaussians

$$p_i(z) = p_i(z; r_i, \sigma_i) = \frac{1}{\sqrt{2\pi\sigma_i^2}} \exp\left(-\frac{(z - r_i)^2}{2\sigma_i^2}\right), \tag{4}$$

where $r_i$ is the mean position of $i^{th}$ component and $\sigma_i$ is the width of its distribution. For their volumetric distributions along the normal holds the formula:

$$v_i(z) = \frac{p_i(z)}{A} V_i. \tag{5}$$

The region of the whole hydrocarbon chains (carbonyl groups excluded) is represented by the so-called *plateau*-function

$$p_i(z; r_{1i}, \sigma_{1i}, r_{2i}, \sigma_{2i}) = 0.5\left(\text{erf}\left(\frac{z - r_{1i}}{\sqrt{2}\sigma_{1i}}\right) - \text{erf}\left(\frac{z - r_{2i}}{\sqrt{2}\sigma_{2i}}\right)\right), \tag{6}$$

$$\text{erf}(x) = \frac{1}{\pi} \int_0^x e^{-t^2} dt,$$

where *r*'s denote the mean borders of hydrocarbon chains of a single lipid molecule (a single bilayer leaflet) and *σ*'s are their corresponding widths. From the widely accepted concept of an ideally filled bilayer hydrophobic region it follows that

$$v_{\text{CH}_2}(z) = p_{\text{CH}_2}(z). \tag{7}$$

To incorporate the presence of water molecules even inside a single leaflet an additional parameter is introduced into the model in contrast to Wiener and White (1992) and Kučerka et al. (2008): the rate of water in a bilayer leaflet, indexed by *a* and called further as *hydration* $E_a$. Then the SLD profile of a single leaflet on the normal is given by

$$\rho_{norm}(z) = \rho_W + \\ + \sum_a (1 - E_a) \left( \sum_i v_{i,a}(z)(\rho_i - \rho_W) + v_{\text{CH}_3,a}(z)(\rho_{\text{CH}_3} - \rho_{\text{CH}_2}) \right), \tag{8}$$

where *a* labels present bilayer leaflets and *i* component groups in leaflet *a*. The SLD profile of a bilayer is obtained by adding two single leaflets together. To circumvent the presence of voids in the hydrocarbon centre of a model bilayer, the corresponding positions and roughnesses of methylene regions in the bilayer centre are set equal. The methyl groups located in the bilayer center and belonging to different leaflets can be described by a common probabilistic and volumetric distribution. To incorporate the effect of different hydration of leaflets $(E_1, E_2)$ within a bilayer, the volumetric distribution of methyl groups is multiplied by a factor $\big((1 - E_1) + (1 - E_2)\big)/2$, in contrast to other components.

The molecular volumes of headgroup components and their distribution widths were taken as average values from the results of molecular dynamics simulations carried out by Klauda *et al.* (2006) for different temperatures in the range 55 – 65 °C. This was allowed by a very small isobaric temperature expansivity coefficient of a PC headgroup (Uhríková et al., 2007). The component volumes of a methylene group $V_{CH_2}$ at both measurement temperatures were from (Uhríková et al., 2007). They are used as internal model parameters and their values are given in Table 1. For the volume of methyl groups on the border of two primitive cells $V_{CH_3}$ we used the fact that the volume of a single methyl group is cca double the size of the volume of a single methylene group $V_{CH_2}$, hence $V_{CH_3} = 8V_{CH_2}$ (Marsh, 2013).

With the knowledge of hydrocarbon chain volumes $V_C$ and the thickness of a complete hydrophobic core of a bilayer $d_C$ one can directly connect the areas per lipid of leaflets in an asymmetrical bilayer

$$A_2 = A_1 \frac{V_C}{A_1 d_C - V_C} \tag{9}$$

and decrease the number of free parameters, or, in the case of a symmetrical bilayer ($A_1 = A_2$), even determine its value $A = 2\,V_C/d_C$ and completely exclude it from the system parameters. However, we found that a better practice is to artificially set the values of $A$'s to an expected values during the beginning of the fitting process and to release them or estimate them later, to avoid the strong influence of other parameters on their values.

A silicon block with a $SiO_2$ layer is incorporated into the model together with water as a part of an environment/matrix, into which a single bilayer or two bilayers are placed the same way as in (8). The Si/$SiO_2$ interface is represented by an error function at $z = 0$ and the $SiO_2$/water interface is represented by an error function with a variable position defining the thickness of an $SiO_2$ layer, thus the SLD profile of a background is

$$\rho_{env} = \rho_{Si} + 0.5\left(1 + \text{erf}\left(\frac{z}{\sqrt{2}\sigma_{Si}}\right)\right)(\rho_{SiO_2} - \rho_{Si}) +$$
$$+ 0.5\left(1 + \text{erf}\left(\frac{z - r_{SiO_2}}{\sqrt{2}\sigma_{SiO_2}}\right)\right)(\rho_W - \rho_{SiO_2}), \tag{10}$$

where $\sigma_{Si}$ represents the roughness of an Si/SiO$_2$ interface, $r_{SiO_2}$ and $\sigma_{SiO_2}$ are the position and the roughness of an SiO$_2$/water interface, respectively, and $\rho$'s are the SLDs of individual parts of the environment. Hence, the complete, updated SLD profile (8) on the normal is

$$\rho_{norm}(z) = \rho_{env} +$$
$$+ \sum_j \left(\sum_{a \text{ in } j} (1 - E_a)\left[\sum_{i \text{ in } a} v_i(z)(\rho_i - \rho_W) + \frac{v_{CH_3,j}(z)}{2}(\rho_{CH_3} - \rho_{CH_2})\right]\right), \tag{11}$$

where $j$ labels bilayers, $a$ leaflets within $j^{th}$ bilayer and $i$ labels component groups and regions within $a^{th}$ leaflet. This way a complete system of a single bilayer or the system of a floating bilayer is modelled in the framework of our model and is transformed via Parratt's recursion (2-5) into a model reflectivity curve.

The model was applied to experimental reflectivity curves by minimization of

$$\chi^2 = \sum_{\substack{\text{exp.}\\\text{points}}} \frac{\left(R_{model}(q_i) - R_{exp}(q_i)\right)^2}{\sigma_i^2} \tag{12}$$

The resolution for each individual point was also included into the modelled reflectivity by convolution of $R(q)$ with the resolution function of the reflectometer $\Delta q(q)$.

*Surface defects evaluation*

A big difference between the modeling of vesicle bilayers and of absorbed bilayers is related to the presence of surface defects for the latter. Surface defects are represented in our model by the presence of hydration water in the hydrophobic core of the bilayer. Presence of water in this region strongly affects the values of other free parameters due to the significant space occupied by hydrocarbon chains. And being the SDP model based on the use of molecular components volume. For this reason we devoted special attention to its evaluation during fitting process. $E_a$ values were updated repeatedly by fitting all three scattering curves after each change of hydrocarbon region borders. After reaching a local minimum and searching in its vicinity for a better one, we approached to a model modification with asymmetrical leaflets hydration. Thus the extent of surface defects can be obtained individually per leaflet with a high relative precision. The previous step has high importance, especially for supported bilayers, where each leaflet undergoes different environment interactions.

## 3. Results and Discussion

*3.1 Supported bilayer*

In the first step, the model was applied to the reflectivity curves of a supported bilayer consisting of diC22:0PC in three different contrasts, as commonly done for reflectometry studies in lipid bilayer systems. The experimental reflectivity curves are shown in Fig. 2A together with the fit curves. Fig. 2B shows the SLD profiles derived from the model as a function of the distance along the bilayer normal. All three reflectivity curves display a minimum between 0.12 and 0.15 Å$^{-1}$, which was found to be sensitive to the thickness of the supported diC22:0PC bilayer as expected due to the interference of reflected waves from the top and the bottom of the bilayer.

The fitting process was divided into three parts. In the first step we kept the relative distance of each component group in polar heads from the hydrocarbon core constant and the same in both leaflets. The area per lipid was also kept at the value 60 Å$^2$, which is slightly higher than usual areas of lipids in the gel ($L_\beta$) phase (Marsh, 2013). Hence, only dominant regions were fitted – the SiO$_2$ layer and the hydrocarbon core sizes with the hydration of the bilayer. After the achievement of a local minimum, when further (realistic) fit improvement was not possible, different hydrations of bilayer leaflets were allowed and the positions of component groups in headgroups were fitted under the condition of their symmetry with regard to hydrocarbon core borders. In the final step, area per lipid $A$ in a combination with leaflet hydrations and individual asymmetrical component group positions were fitted repeatedly.

To evaluate the goodness of the fit curves of the spread of normalized residuals $\left(R_{model}(q_i) - R_{exp}(q_i)\right)/\sigma_i$ values was monitored. For the present case it always lies in the interval (-2, 2) almost for all $q$-values.

In contrast to König et al. (1996), Charitat et al. (1999) and Gutberlet et al. (2004), where adsorbed PC bilayers were measured by SNR and Stidder et al. (2007), where SNR was applied on supported dipalmitoylphosphatidylethanolamine (diC16:0PE) bilayers, we did not detect a bulk water layer between the SiO$_2$ surface layer and the adsorbed diC22:0PC bilayer. From the position of a choline group in the bilayer leaflet closer to the SiO$_2$ layer, we can deduce that the adsorbed bilayer is in a direct contact with the hydrated silicon oxide layer. This is not surprising and in an agreement with other SNR results (Gerelli et al., 2012, 2013).

The structural parameters obtained by the analysis of diC22:0PC bilayer according to the SDP model are summarized in Table 1.

The thickness of the hydrocarbon region is 44.9 ± 0.1 Å. If we suppose that the chain tilt of lipid bilayers in the gel phase does not depend on their length (Marsh, 2013), then using the value of the hydrocarbon thickness of, for example, diC14:0PC in the gel phase $2D_C$ = 30.3 Å (Tristram-Nagle et al., 2002), we can estimate $2D_C$ of a non-perturbed diC22:0PC bilayer to $2D_C$ = 47.6 Å. Our value $2D_C$ = 44.9 ± 0.1 Å is lower, what can be explained by the interaction with the $SiO_2$ surface. On the other side, our $2D_C$ value is in a very good agreement with a similar estimation ($2D_C$ = 44.0 Å) if the results of Charitat et al. (1999) for diC16:0PC and diC18:0PC are used as reference..

The area per lipid of the leaflet in contact with the $SiO_2$ layer $A$ = 55.0 ± 1.0 Å$^2$ is slightly higher, but within the experimental accuracy, than in the outer leaflet, which is in contact with bulk water, A = 52.9 ± 1.2 Å$^2$. This might be caused by a different kind of environment interactions or by a different type of deposition (the lower leaflet was deposited by a vertical deposition and the outer one by horizontal one). This corresponds to the results of Charitat et al. (1999) and Gutberlet et al. (2004), where non-symmetrical box models were applied.

We found that surface defects were present in the supported bilayers. As mentioned before, the extent of defects is evaluated by the amount of water detected in the hydrophobic core of the bilayer. The overall coverage of the silicon block, defined as a complement to unit hydration, was ca 90 % (89.5 ± 0.5 % and 91.2 ± 0.4 % for inner and outer leaflet respectively) (Table 1). The hydration of the lower leaflet core is slightly higher than the hydration of the outer one, but the difference is less than 2 %

*3.2 Floating bilayer*

A floating diC16:0PC bilayer over a supported diC22:0PC bilayer was measured in three different contrasts at 25 °C and 55 °C. The fitting process for the floating bilayer was similar to the procedure for the supported bilayer described above, but it was considered as completely symmetrical

(component group positions, hydration, hydrocarbon core roughnesses). For the analysis of the diC22:0PC bilayer, used as support in this sample, In the case of the supported bilayer we varied only its relative position to the $SiO_2$ surface and its hydration keeping its structure the same as obtained from the adsorbed diC22:0PC bilayer described in the previous section.

The best obtained model fit and its SLD profiles are shown in Fig. 3A and 3B. We tried to fit the floating bilayer as asymmetrical or increase its roughness to include any effects coming from different depositions of its leaflets, but it did not improve the fit significantly. The hydrocarbon region thickness $2D_C = 35.9 \pm 0.5$ Å is higher than hydrocarbon region thickness of the same system at the same phase $2D_C = 34$ Å obtained by Charitat et al. (1999) and $2D_C = 32.0 \pm 2$ Å obtained by Fragneto et al. (2003) indicating that structural differences were present.

At 55 °C we could obtain a much better global fit. The fit curves with the corresponding SLD profiles are depicted on Fig. 4A and 4B. A change in the hydrocarbon region thickness, with respect to the value found at the lower temperature, was observed. This was expected since the gel-to-liquid phase transition temperature for d62-diC16:0PC is $T_m = 39$°C. The hydrocarbon thickness we obtained $2D_C = 27.2 \pm 0.5$ Å is in a very good agreement with the thickness $2D_C = 27.9$ Å obtained by Kučerka et al. (2011b) with a similar lipid bilayer model from small-angle neutron scattering on unilamellar vesicles. However, the region roughness is in our case evidently higher $\sigma_{CH_2} = 4.8 \pm 0.5$ Å than the value of Kučerka et al. (2011b) $\sigma_{CH_2} = 2.5$ Å. This is probably caused by fluctuations arising from higher freedom of the floating bilayer in comparison to the bilayer in the form of a vesicle.

If we compare the estimated hydration of the floating bilayer at 25 °C and its value at 55 °C, we can see that it was decreased by more than 15 %. This leads to the well known conclusion that the fluid phase is more suitable for the reorganization of lipid molecules into well-ordered bilayer than the gel phase.

If we assign $r_{CHO,2} + \sigma_{CHO}$ and $r_{CHO,1} - \sigma_{CHO}$ as the outer border of the supported bilayer and the lower border of the floating bilayer, respectively, we can estimate the interbilayer water thickness $D_W$. Between its estimated value from the data at 25 °C $D_W = 22.4$ Å and the value obtained from the data at 55 °C $D_W = 20.5$ Å is the difference of ca. 2 Å. Albeit the validity of this approximation is strongly dependent on which experimental technique is applied, the tendency agrees with a formerly observed behaviour of diC16:0PC bilayers across their main phase transition (Fragneto et al., 2012).

The interface area per lipid molecule $A = 66.6 \pm 0.2$ Å$^2$ in the floating bilayer at 55 °C is slightly higher than its value obtained by Kučerka et al. (2011b), but the difference in the number of molecules intercalated into the hydrophilic bilayer region per lipid molecule is less than 1 water molecule. If we use the same estimation for the bilayer borders as it was already mentioned above and the total phosphatidylcholine headgroup volume, given as the sum of volumes of its components, we can estimate the average number of water molecules in the floating bilayer hydrophilic region per one lipid molecule to $N_W = 14.4$.

The arrangement of component groups in the hydrophilic regions differs slightly from the arrangement obtained by Kučerka et al. (2011b). Firstly, the difference between the positions of the carbonyl-glycerol group is only 0.5 Å and in our case it is shifted from the hydrocarbon core. Similar shifts can be seen also for the phosphate and the choline groups. As the shift increases with the distance from the hydrocarbon core, we assume that they are predominantly caused by the bilayer fluctuations. There is also the effect of different component specifications in the case of phosphate and choline groups, when one compares our bilayer model with the model of Kučerka et al. (2008, 2011b), but as it can be seen in (Heberle et al., 2012), the shift is around 1 Å. Hence we consider bilayer fluctuations as the main reason of a wider spread of headgroup components on the bilayer normal.

## 4. Conclusions

The modified SLD profile model was applied successfully for the first time for the analysis of neutron reflectivity from both supported and floating bilayers. It was found that its application can be negatively influenced by disorder mainly in the case of a floating bilayer. This effect was suppressed by heating the bilayer into the liquid phase. In both cases, for a supported bilayer as well as a floating bilayer in the fluid phase, the SLD profile model revealed their internal structure in terms of headgroup components.

In the case of a supported diC22:0PC bilayer results showed a slightly asymmetrical structure as in the hydration of its leaflets, the leaflet interface areas per lipid molecule and the structure of its hydrophilic regions.

For the floating d62-diC16:0PC bilayer in the gel phase at 25 °C only its full thickness and average hydration were estimated. A more detailed analysis was not possible very likely because of the disorder present in the sample before the thermal annealing.

The reflectivity curves of the floating d62-diC16:0PC bilayer in the fluid phase at 55 °C were successfully simultaneously fitted by the applied model. The results display a symmetrical structure and lower hydration in comparison to its state in the gel phase after deposition. The mean positions of component groups in hydrophilic regions were shifted from the bilayer center increasingly with their relative distances from the bilayer center in comparison to the same bilayer structure obtained from small-angle neutron scattering on bilayers in the form of unilamellar vesicles. Moreover, the roughness of its hydrocarbon region was about twice than the roughness in unilamellar vesicles. On the other side the mean thickness of the hydrocarbon region was in a very good agreement with small-angle neutron measurements. Hence, we conclude that the floating bilayer fluctuates due to its freedom at higher rate than the bilayer in unilamellar vesicles.

These results show that the system is appropriate as biomembrane model. It has been found previously that when both bilayers in a double bilayer system are brought in the fluid phase, there is considerable mixing of the lipids from the two bilayers (Gerelli et al., 2012; Rondelli et al., 2013). By using a longer chain lipid (diC22:0PC) as supporting bilayer this stays the gel phase and there is no mixing with the floating one. The system represents therefore a step forward the preparation of complex model membrane systems for structural studies of the floating bilayer.

# Acknowledgements

This work was supported by the VEGA 1/0159/11 grant. The authors would like to acknowledge ILL for beamtime and for the support provided during the measurements. We acknowledge the use of the preparation and characterization tools provided by the Partnership for Soft Condensed Matter (PSCM) at ILL. MB also thanks the Central European Neutron Initiative for a one-year scholarship and the staff of Large Scales Structure Group in Institute Laue-Langevin for help, discussions and hospitality.

# References

Born, M., Wolf, E., 1999. Principles of optics: electromagnetic theory of propagation, interference and diffraction of light. Cambridge University Press, Cambridge.

Charitat, T., Bellet-Amalric, E., Fragneto, G., Graner, F., 1999. Adsorbed and free lipid bilayers at the solid-liquid interface. Eur. Phys. J. B 8, 583–593. doi:10.1007/s100510050725

Cubitt, R., Fragneto, G., 2002. D17: the new reflectometer at the ILL. Appl. Phys. A 74, 329–331. doi:10.1007/s003390201611

Daillant, J., Bellet-Amalric, E., Braslau, A., Charitat, T., Fragneto, G., Graner, F., Mora, S., Rieutord, F., Stidder, B., 2005. Structure and fluctuations of a single floating lipid bilayer. Proc. Natl. Acad. Sci. U. S. A. 102, 11639–11644. doi:10.1073/pnas.0504588102

Fragneto, G., Charitat, T., Bellet-Amalric, E., Cubitt, R., Graner, F., 2003. Swelling of phospholipid floating bilayers: the effect of chain length. Langmuir 19, 7695–7702. doi:10.1021/la026972x

Fragneto, G., Charitat, T., Daillant, J., 2012. Floating lipid bilayers: models for physics and biology. Eur. Biophys. J. 41, 863–874. doi:10.1007/s00249-012-0834-4

Fragneto, G., Charitat, T., Graner, F., Mecke, K., Perino-Gallice, L., Bellet-Amalric, E., 2001. A fluid floating bilayer. Europhys. Lett. 53, 100–106. doi:10.1209/epl/i2001-00129-8

Gerelli, Y., Porcar, L., Fragneto, G., 2012. Lipid rearrangement in DSPC/DMPC bilayers: A neutron reflectometry study. Langmuir 28, 15922–15928. doi:10.1021/la303662e

Gerelli, Y., Porcar, L., Lombardi, L., Fragneto, G., 2013. Lipid exchange and Flip-Flop in solid supported bilayers. Langmuir 29, 12762–12769. doi:10.1021/la402708u

Gutberlet, T., Steitz, R., Fragneto, G., Klösgen, B., 2004. Phospholipid bilayer formation at a bare Si surface: a time-resolved neutron reflectivity study. J. Phys. Condens. Matter 16, S2469. doi:10.1088/0953-8984/16/26/020

Heberle, F.A., Pan, J., Standaert, R.F., Drazba, P., Kučerka, N., Katsaras, J., 2012. Model-based approaches for the determination of lipid bilayer structure from small-angle neutron and X-ray scattering data. Eur. Biophys. J. 41, 875–890. doi:10.1007/s00249-012-0817-5

Heimburg, T., 2007. Thermal Biophysics of Membranes. Wiley-VCH Verlag, Weinheim.

Heller, W.T., 2010. Small-angle neutron scattering and contrast variation: a powerful combination for studying biological structures. Acta Crystallogr. D 66, 1213–1217. doi:10.1107/S0907444910017658

Hope, M.J., Bally, M.B., Mayer, L.D., Janoff, A.S., Cullis, P.R., 1986. Generation of multilamellar and unilamellar phospholipid vesicles. Chem. Phys. Lipids 40, 89–107. doi:10.1016/0009-3084(86)90065-4

Hughes, A.V., Howse, J.R., Dabkowska, A., Jones, R.A.L., Lawrence, M.J., Roser, S.J., 2008. Floating lipid bilayers deposited on chemically grafted phosphatidylcholine surfaces. Langmuir 24, 1989–1999. doi:10.1021/la702050b

Katsaras, J., Yang, D.S., Epand, R.M., 1992. Fatty-acid chain tilt angles and directions in dipalmitoyl phosphatidylcholine bilayers. Biophys. J. 63, 1170–1175. doi:10.1016/S0006-3495(92)81680-6

Klauda, J.B., Kučerka, N., Brooks, B.R., Pastor, R.W., Nagle, J.F., 2006. Simulation-based methods for interpreting X-ray data from lipid bilayers. Biophys. J. 90, 2796–2807. doi:10.1529/biophysj.105.075697

König, B.W., Krueger, S., Orts, W.J., Majkrzak, C.F., Berk, N.F., Silverton, J.V., Gawrisch, K., 1996. Neutron reflectivity and atomic force microscopy studies of a lipid bilayer in water adsorbed to the surface of a silicon single crystal. Langmuir 12, 1343–1350. doi:10.1021/la950580r

Kučerka, N., Gallová, J., Uhríková, D., Balgavý, P., Bulacu, M., Marrink, S.-J., Katsaras, J., 2009. Areas of monounsaturated diacylphosphatidylcholines. Biophys. J. 97, 1926–1932. doi:10.1016/j.bpj.2009.06.050

Kučerka, N., Holland, B.W., Gray, C.G., Tomberli, B., Katsaras, J., 2011a. Scattering density profile model of POPG bilayers as determined by molecular dynamics simulations and small-angle neutron and X-ray scattering experiments. J. Phys. Chem. B 116, 232–239. doi:10.1021/jp208920h

Kučerka, N., Nagle, J.F., Sachs, J.N., Feller, S.E., Pencer, J., Jackson, A., Katsaras, J., 2008. Lipid bilayer structure determined by the simultaneous analysis of neutron and X-ray scattering data. Biophys. J. 95, 2356–2367. doi:10.1529/biophysj.108.132662

Kučerka, N., Nieh, M.-P., Katsaras, J., 2011b. Fluid phase lipid areas and bilayer thicknesses of commonly used phosphatidylcholines as a function of temperature. Biochim. Biophys. Acta 1808, 2761–2771. doi:10.1016/j.bbamem.2011.07.022

MacDonald, R.C., MacDonald, R.I., Menco, B.P.M., Takeshita, K., Subbarao, N.K., Hu, L., 1991. Small-volume extrusion apparatus for preparation of large, unilamellar vesicles. Biochim. Biophys. Acta 1061, 297–303. doi:10.1016/0005-2736(91)90295-J

Marsh, D., 2013. Handbook of Lipid Bilayers, 2nd ed. CRC Press, Boca Raton.

Nieh, M.-P., Ku\vcerka, N., Katsaras, J., 2009. Spontaneously formed unilamellar vesicles. Meth. Enzymol. 465, 3–20. doi:10.1016/S0076-6879(09)65001-1

Parratt, L.G., 1954. Surface studies of solids by total reflection of X-rays. Phys. Rev. 95, 359–369. doi:10.1103/PhysRev.95.359

Richard, D., Ferrand, M., Kearley, G.J., 1996. Analysis and visualisation of neutron-scattering data. J. Neutron Res. 4, 33–39. doi:10.1080/10238169608200065

Rondelli, V., Del Favero, E., Motta, S., Cantù, L., Fragneto, G., Brocca, P., 2013. Neutrons for rafts, rafts for neutrons. European Physical Journal E 36. doi:10.1140/epje/i2013-13073-4

Sackmann, E., 1996. Supported membranes: scientific and practical applications. Science 271, 43–48. doi:10.1126/science.271.5245.43

Shekhar, P., Nanda, H., Lösche, M., Heinrich, F., 2011. Continuous distribution model for the investigation of complex molecular architectures near interfaces with scattering techniques. Journal Of Applied Physics 110, 102216.

Stidder, B., Fragneto, G., Roser, S.J., 2007. Structure and stability of DPPE planar bilayers. Soft Matter 3, 214–222. doi:10.1039/b612538g

Tristram-Nagle, S., Liu, Y., Legleiter, J., Nagle, J.F., 2002. Structure of gel phase DMPC determined by X-ray diffraction. Biophys. J. 83, 3324–3335. doi:10.1016/S0006-3495(02)75333-2

Uhríková, D., Rybár, P., Hianik, T., Balgavý, P., 2007. Component volumes of unsaturated phosphatidylcholines in fluid bilayers: a densitometric study. Chem. Phys. Lipids 145, 97–105. doi:10.1016/j.chemphyslip.2006.11.004

Vacklin, H.P., Tiberg, F., Fragneto, G., Thomas, R.K., 2005. Composition of supported model membranes determined by neutron reflection. Langmuir 21, 2827–2837. doi:10.1021/la047389e

Wiener, M.C., White, S.H., 1992. Structure of a fluid dioleoylphosphatidylcholine bilayer determined by joint refinement of x-ray and neutron diffraction data. III. Complete structure. Biophys. J. 61, 434–447. doi:10.1016/S0006-3495(92)81849-0

# Tables

| Internal parameter | Value |
|---|---|
| $\rho_{Si}$ | $2.07 \times 10^{-6}$ Å$^{-2}$ |
| $\rho_{SiO_2}$ | $3.41 \times 10^{-6}$ Å$^{-2}$ |
| $V_{CHO}^{a)}$ | 108.1 Å$^3$ |
| $\sigma_{CHO}^{a)}$ | 2.76 Å |
| $V_{PO}^{a)}$ | 69.2 Å$^3$ |
| $\sigma_{PO}^{a)}$ | 2.31 Å |
| $V_{GC}^{a)}$ | 147.3 Å$^3$ |
| $\sigma_{GC}^{a)}$ | 2.37 Å |
| $V_{CH_2}^{b)}$ | 27.5 Å$^3$ (25 °C)<br>28.5 Å$^3$ (55 °C) |
| $\sigma_{CH_2}^{b)}$ | $2V_{CH_2}$ |

**Table 1:** The internal parameters and their values used in the SDP model of a bilayer. [a)] Obtained by Klauda et al. (2006). [b)] Obtained by Uhríková et al. (2007). [c)] References in (Marsh, 2013).

| Free parameter | diC22:0PC | d62-diC16:0PC at 25 °C | d62-diC16:0PC at 55 °C | diC16:0PC** at 66 °C |
|---|---|---|---|---|
| $\sigma_{Si}$ [Å] | 4.2 ± 0.3 | 4.1 ± 0.2 | 4.1 ± 0.2 | - |
| $r_{SiO_2/H_2O}$ [Å] | 14.0 ± 0.1 | 13.7 ± 0.1 | 13.7 ± 0.1 | - |
| $\sigma_{SiO_2}$ [Å] | 3.0 ± 0.5 | 3.7 ± 0.2 | 3.7 ± 0.2 | - |
| $A$ [Å²] | 55.5 ± 1.0 <br> 52.9 ± 1.2 | 64.4 ± 0.3 | 66.6 ± 0.2 | 65.0 |
| $E_i$ | 0.105 ± 0.005 <br> 0.088 ± 0.004 | 0.443 ± 0.001 <br> (0.171 ± 0.001) | 0.291 ± 0.001 <br> (0.159 ± 0.004) | - |
| $r_{CHO,1}$ [Å] | 18.7 ± 0.6 | 103.3* | 101.4 ± 0.4 | 104.3 |
| $r_{PO,1}$ [Å] | 20.2 ± 1.0 | 106.8* | 104.3 ± 0.7 | 105.1 |
| $r_{GC,1}$ [Å] | 23.0 ± 0.5 | 113.6* | 108.7 ± 0.4 | 109.3 |
| $r_{CH_2,1}$ [Å] | 25.3 ± 0.5 | 115.2 ± 0.5 | 110.0 ± 0.5 | 110.0 |
| $\sigma_{CH_2,1}$ [Å] | 2.8 ± 0.6 | 2.9 ± 0.5 | 4.8 ± 0.5 | 2.5 |
| $r_{CH_2,2}$ [Å] | 70.2 ± 0.5 | 151.1 ± 0.5 | 137.2 ± 0.5 | 137.9 |
| $\sigma_{CH_2,2}$ [Å] | 2.2 ± 0.6 | 2.9 ± 0.5 | 4.8 ± 0.5 | 2.5 |
| $r_{GC,2}$ [Å] | 71.3 ± 0.5 | 152.7* | 138.4 ± 0.4 | 138.7 |
| $r_{PO,2}$ [Å] | 74.8 ± 1.6 | 159.4* | 142.8 ± 0.7 | 142.9 |
| $r_{CHO,2}$ [Å] | 75.4 ± 0.4 | 163.0* | 145.7 ± 0.4 | 143.7 |

**Table 2:** The best obtained values of the bilayer model structure parameters for a supported diC22:0PC bilayer and a floating d62-diC16:0PC bilayer at 25 °C and 55 °C. For diC22:0PC area per lipid the first and the second number represent the area per lipid in the first and in the second leaflet in *z*-direction, respectively. The same applies for hydrations $E_i$. The asterisk (*) denotes the position values obtained with fixed relative positions of component groups and equal to ones obtained by Kučerka et al. (2011b). For floating bilayers the hydration in parentheses denotes the average hydration of the supported bilayer. The errors were obtained by cyclic fitting refinements of parameters in groups of two or three. The double asterisk (**) in the last column denotes recalculated diC16:0PC bilayer structure parameters obtained by Kučerka et al. (2011b) from SANS on unilamellar vesicles are shown. The structure parameters are recalculated for better visualization to the same position of the lower hydrocarbon region border of a floating bilayer ($r_{CH_2,1} = 110$ Å).

# Figure captions

**Fig. 1:** Scheme of component group volumetric probabilities inside of the modified SLD profile model of a supported and a floating bilayers. The first and second numbers in the group names label the leaflet and the bilayer respectively.

**Fig. 2:** (A) The normalized reflectivity curves of a supported diC22:0PC bilayer on the hydrophilic surface of a silicon substrate. The bilayer was measured in three different contrasts: pure $H_2O$ (squares, blue line), mixture of $H_2O/D_2O$ (triangles, green line) and in pure $D_2O$ (circles, red line). The lines represent the best simultaneous fit of the component bilayer model to the data. (B) The corresponding SLD profiles. The lines represent the bilayer model in three different contrasts and correspond to the fitting lines.

**Fig. 3:** (A) The normalized reflectivity curves of a floating d62-diC16:0PC bilayer over a diC22:0PC bilayer at 25 °C in three different contrasts. The lines represent the best simultaneous fit of the component bilayer model to the data. (B) The corresponding SLD profiles.

**Fig. 4:** (A) The normalized reflectivity curves of a floating d62-diC16:0PC bilayer over a diC22:0PC bilayer at 55 °C in three different contrasts. The lines represent the best simultaneous fit of the component bilayer model to the data. (B) The corresponding SLD profiles.

# Figures

**Fig. 1**

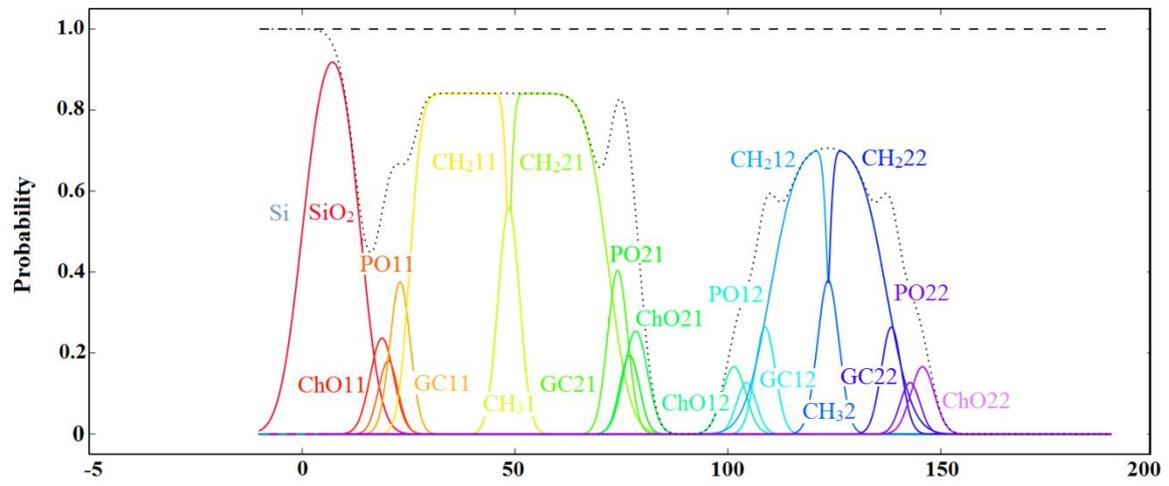

**Fig. 2**

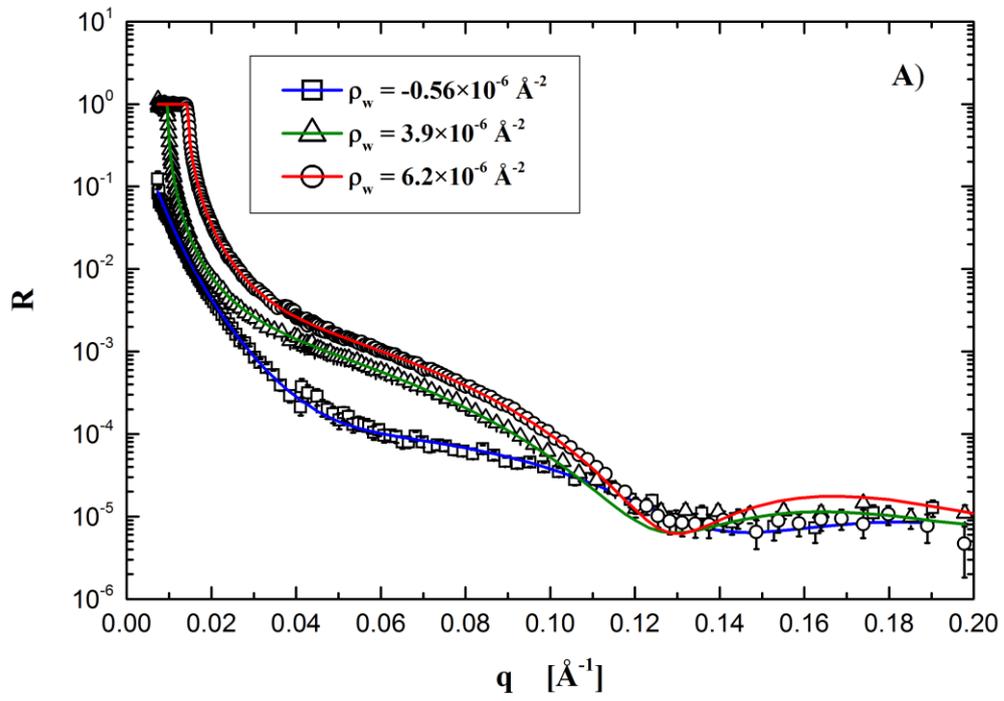

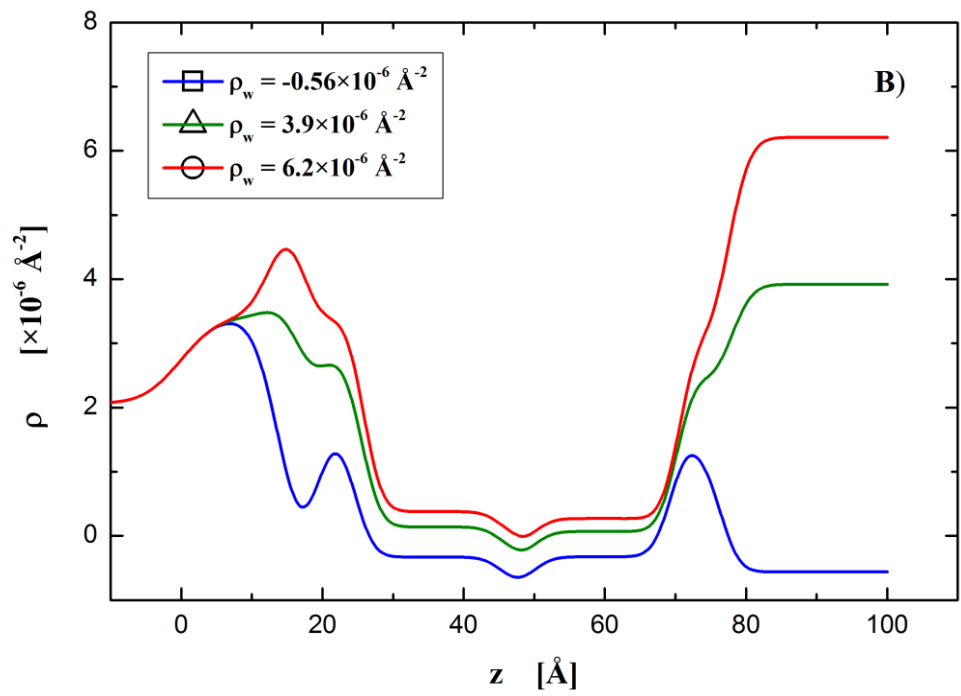

**Fig. 3**

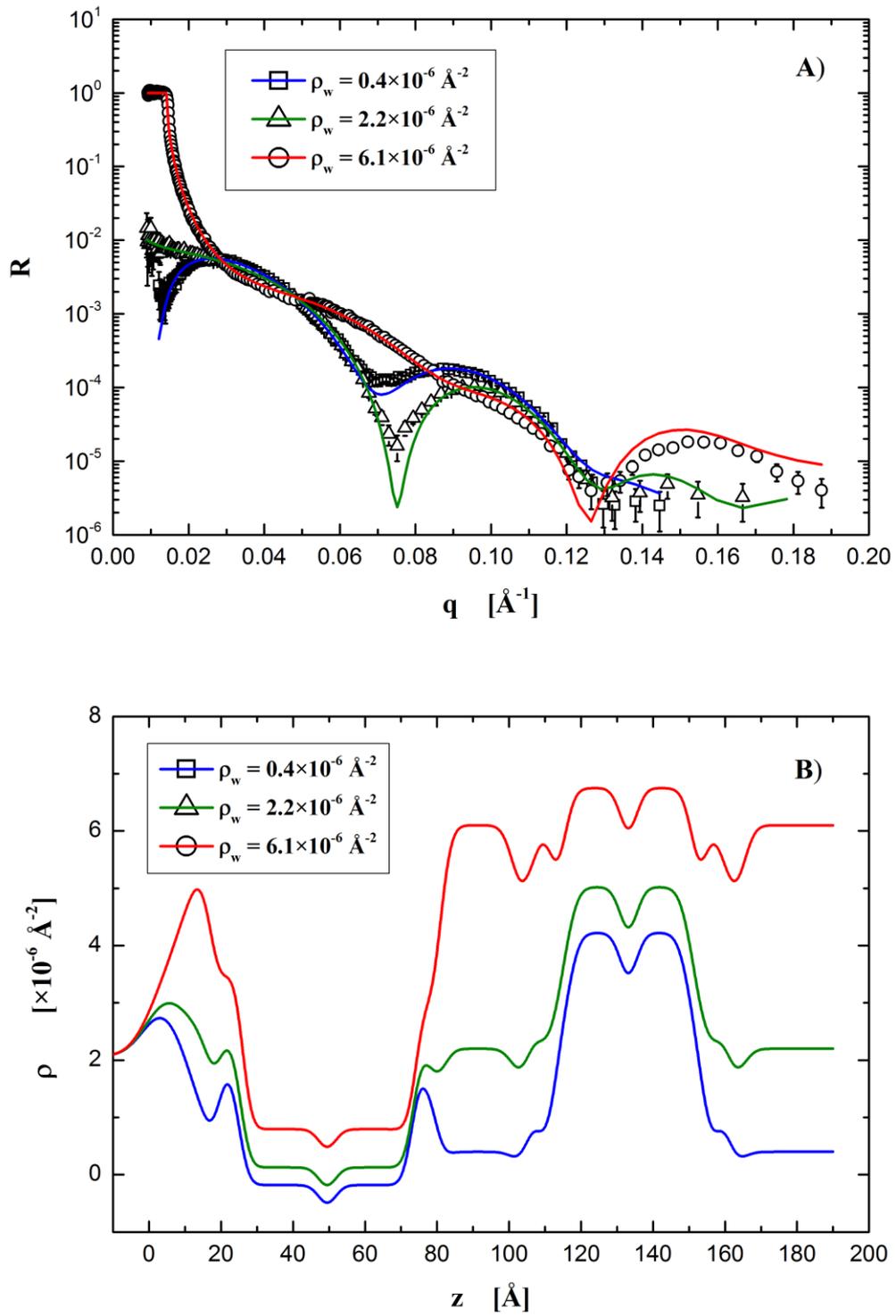

**Fig. 4**

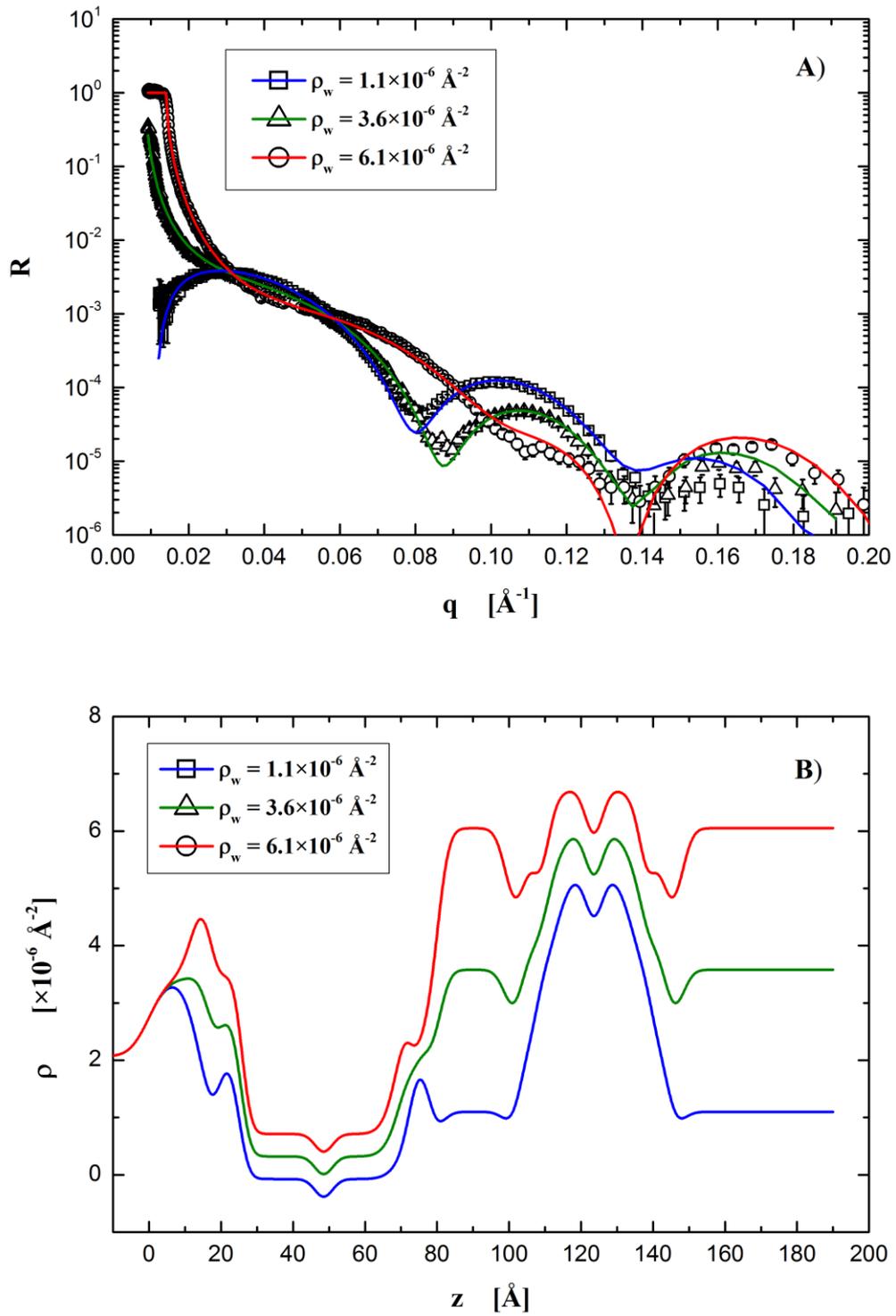